\documentclass{vldb-nonres}
\usepackage{color}
\usepackage{url}
\usepackage{hyperref}
\usepackage{xspace}
\usepackage{pslatex}
\usepackage{enumitem}
\usepackage{epstopdf}

\usepackage{amsmath}
\usepackage{amsfonts}
\usepackage{euscript}
\usepackage{graphicx}
\usepackage{tabularx}
\usepackage{tabulary}
\usepackage{longtable}
\usepackage{balance}  
\usepackage{microtype}

\newcommand{\pbq}{\textsc{PackageBuilder}\xspace}
\newcommand{\pql}{PaQL\xspace}

\begin{document}
    
    \makeatletter
    \def\@copyrightspace{\relax}
    \makeatother

\title{PackageBuilder: From Tuples to Packages}

\numberofauthors{1}
\author{
\alignauthor
\phantom{${}^{\star}$}Matteo Brucato${}^{\star}$ \hfill \phantom{${}^{\star}$}Rahul Ramakrishna${}^{\star}$ \hfill  \phantom{${}^{\S}$}Azza Abouzied${}^{\S}$ \hfill
\phantom{${}^{\star}$}Alexandra Meliou${}^{\star}$\\
\vspace{2mm}
\begin{tabular}{cccc}
\affaddr{${}^{\star}$School of Computer Science\phantom{${}^{\star}$}} & & &  
\affaddr{${}^{\S}$Computer Science\phantom{${}^{\S}$}} \\
\affaddr{University of Massachusetts} & & & \affaddr{New York University} \\
\affaddr{Amherst, USA} & & & \affaddr{Abu Dhabi, UAE} \\
\affaddr{\eaddfnt{\{matteo,rahulram,ameli\}@cs.umass.edu}} & & & \affaddr{\eaddfnt{azza@nyu.edu}} \\
\end{tabular}
}

\date{3}
\maketitle

\begin{abstract}
	In this demo, we present \pbq, a system that extends database systems to 
	support package queries. A package is a collection of tuples that 
	individually satisfy base constraints and collectively satisfy global 
	constraints.
	The need for package support arises in a variety of scenarios: For example,
	in the creation of meal plans, users are not only interested in the nutritional
	content of individual meals (base constraints), but also care to specify 
	daily consumption limits and control the balance of the entire plan (global constraints).
	We introduce \pql, a declarative SQL-based package query language, and the
	interface abstractions which allow users to interactively specify package
	queries and easily navigate through their results. To efficiently
	evaluate queries, the system employs pruning and heuristics, as well as
	state-of-the-art constraint optimization solvers.
	We demonstrate \pbq by allowing attendees to interact with the system's 
	interface, to define \pql queries and to observe how query evaluation is 
	performed.
\end{abstract}

\section{Introduction}\label{sec:intro}
Traditional database queries define constraints (selection predicates) that
each tuple in the result needs to satisfy. Although they are
undoubtedly expressive and powerful, they prove inadequate in scenarios that require
a set of answer tuples to satisfy constraints \emph{collectively}. Such scenarios
arise in a variety of applications:

\begin{description}[leftmargin=0mm, topsep=0mm, itemsep=0mm]
    \item[Meal planner:] An athlete needs to put together a dietary plan in
    preparation for a race. She wants a high-protein set of three gluten-free meals for
    the day, having in total between 2,000 and 2,500 calories.
    It is easy to exclude meals with gluten, as this condition can be
    enforced on each individual meal (tuple) with a regular selection
    predicate.
    The other constraints (e.g., total calories), however, need 
    to be verified collectively over the entire package.
    \item[Vacation planner:] A couple wants to organize a relaxing vacation at
    a tropical destination. They do not want to spend more than \$2,000 on
    flights and hotels combined. They also want to be in walking distance from
    the beach, unless their budget can fit a rental car, in which case
    they are willing to stay farther away. Building the ideal vacation package
    is challenging as the choice of hotels affects the choice of flights and car rentals.
    \item[Investment portfolio:] A broker wants to construct an investment
    portfolio for one of her clients. The client has a budget of \$50K, wants
    to invest at least 30\% of the assets in technology, and wants a balance
    of short-term and long-term options. The broker cannot select each stock
    option individually, but rather needs to find a stock package that
    satisfies all these constraints collectively.
\end{description}

These examples cannot be expressed by traditional SQL queries. We demonstrate
\pbq, a system that augments database functionality to support the creation of
\emph{packages}.
A \emph{package} is a collection of tuples that individually satisfy
\emph{base constraints} and collectively satisfy \emph{global constraints}.
The base constraints are equivalent to regular selection predicates, and can
be evaluated individually for each tuple. For example, in the meal planner
application, the gluten-free restriction is a base constraint, as it can be
verified independently on each meal. In contrast, the requirement that the total 
amount of calories should be between 2,000 and 2,500 cannot be evaluated on each
meal individually, but needs to be assessed over a collection of meals.
\begin{figure*}[t]
\begin{center}
\includegraphics[width=0.93\textwidth]{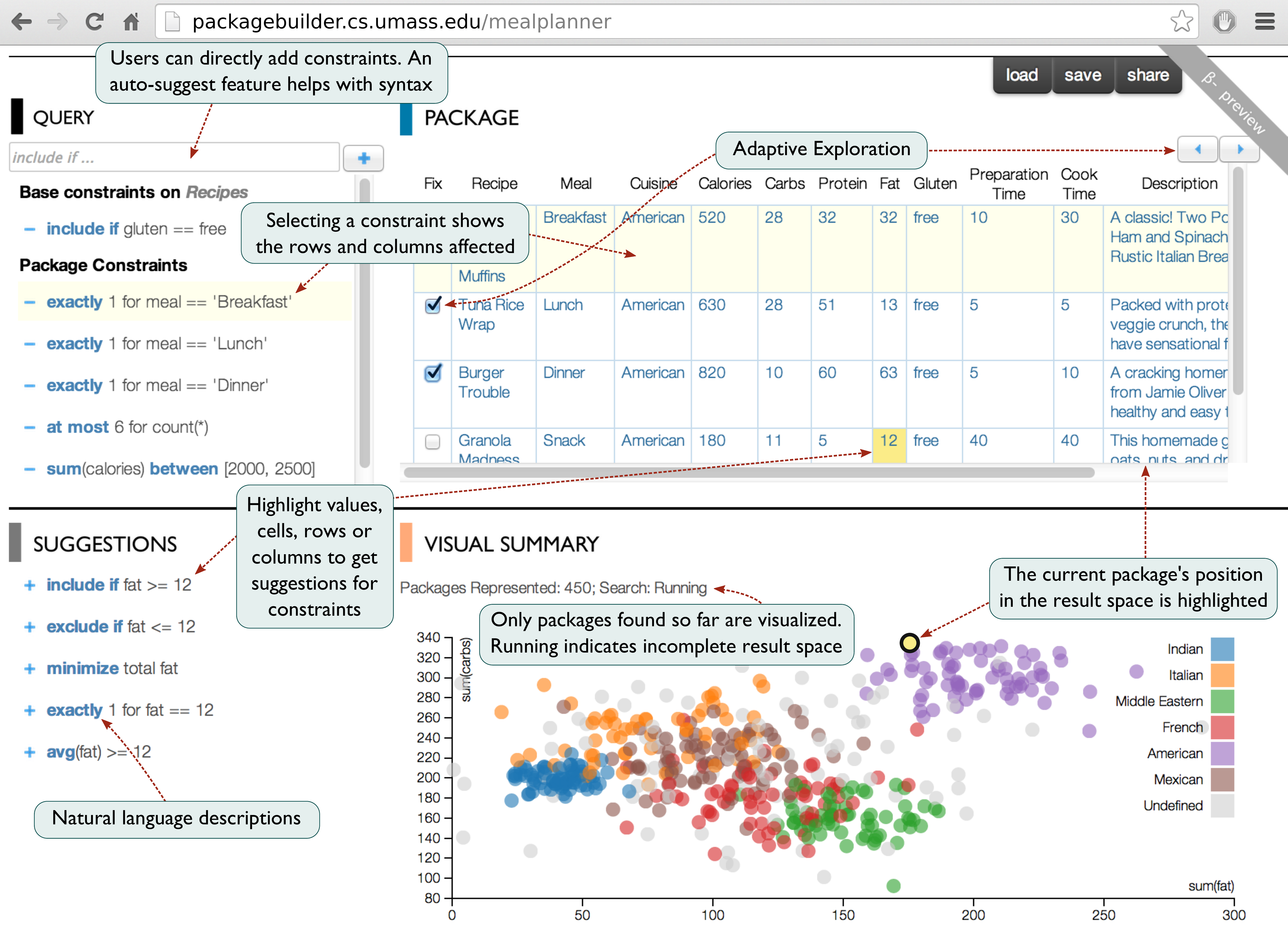}
\caption{The visual interface of \pbq provides different visual representations of packages, and allows the user to interactively manipulate package queries.}
\vspace{-3mm}
\label{fig:template}
\end{center}
\end{figure*}
Our system addresses three main challenges:

\begin{description}[leftmargin=0mm, topsep=0mm, itemsep=0mm]
    \item[Language specification:] Even though many use cases motivate support
    for package queries, this class of queries remains largely unsupported, with few tools
    targeting domain-specific packages (e.g., CourseRank supports building
    course packages~\cite{course-rank}). As part of this demonstration, we
    will present \pql, a declarative query language that supports package
    specifications. \pql is designed with simple extensions to standard SQL.
    Those familiar with SQL should find it intuitive and easy to use
    (Section \ref{sec:language}).
    \item[Interactive specification:] Even traditional SQL queries can often
    be challenging to specify for novice users. To enable user-friendly
    database applications, several systems now employ application-independent
    visual metaphors for SQL query specification~\cite{zloof1977query,
    olsten1998viqing, polaris}. Package queries are fundamentally harder to express and
    evaluate compared to traditional SQL. Therefore, it is increasingly
    important to provide visual paradigms to guide users through query building 
    and through navigating and refining the results.
    \pbq helps users to interactively
    specify base and global constraints for their packages. The system
    interface also allows users to visually navigate through the solution space
    and to easily refine the resulting packages (Section \ref{sec:interface}).
    \item[Evaluation:] In traditional database queries, the size of the answer
    is polynomial in the size of the input data. This is not true for package
    queries: If $n$ tuples satisfy the base constraints of a package query, there
    are $\Omega(2^n)$ candidate packages that can potentially satisfy the user's global
    constraints. This makes the evaluation of package queries particularly challenging.
    With an exponential search space, efficiently searching for packages that
    satisfy the users' constraints requires applying non-trivial pruning
    techniques and search heuristics (Section~\ref{sec:complexity}).
\end{description}

We proceed to describe the main three aspects of our system that are motivated
by these challenges. We conclude with a description of a demonstration
scenario that is illustrative of the system (Section~\ref{sec:scenario}).

\section{\pql: Package Query Language}
\label{sec:language}

Our \pbq system extends traditional database functionality to provide
full-fledged support for packages. We identify two important reasons to
support packages at the database level, rather than at the application
level:
(a) The data used to construct packages typically resides in a database
system, and packages themselves are structured data objects that 
should naturally be stored in and manipulated by a database system.
(b) The features of packages and the algorithms for constructing them are not
unique to each application; therefore, the burden of package support should be
shifted away from the application developers.

We designed \pql, a declarative SQL-based query language for specifying package queries.
The following query builds the athlete's daily meal plan described in Section~\ref{sec:intro}:

{\sf
\begin{tabbing}
\hspace*{5mm}\=\hspace*{2cm}\=\hspace*{3cm}\= \kill
\>SELECT  \>\textbf{PACKAGE}(R) AS P \\
\>FROM    \>Recipes R \\
\>WHERE   \>R.gluten $=$ `free' \\
\>\textbf{SUCH THAT}\>COUNT($*$) $=$ $3$ AND \\
\>        \>SUM(P.calories) BETWEEN $2000$ AND $2500$ \\
\>\textbf{MAXIMIZE}\> SUM(P.protein)
\end{tabbing}}

The introduction of the keyword {\sf PACKAGE} differentiates \pql queries from
traditional SQL queries. Semantically, {\sf PACKAGE} constructs
\emph{multisets} from subsets of tuples from the base relations listed in the {\sf
FROM} clause. With no further constraints, there are infinitely many packages
that can be built from non-empty base relations\footnote{This assumes that a
tuple from an input relation can appear multiple times in the package
result.}. Users can limit the number of times a tuple from the input relation
{\sf R} can appear in the package result by adding a {\sf REPEAT} keyword in
the {\sf FROM} clause. For example, ``{\sf FROM Recipes R REPEAT $k$}'' would
allow a tuple to be repeated up to $k$ times in a package.

A package query defines two types of constraints. \emph{Base constraints},
defined in the {\sf WHERE} clause, are equivalent to selection predicates
and can be evaluated with standard SQL: any tuple in the package needs to
\emph{individually} satisfy all the base constraints. In the example query, the
base constraint ``{\sf R.gluten $=$ `free'}'' specifies that each meal in the
package should be gluten-free. \emph{Global constraints} are defined in the
{\sf SUCH THAT} clause. They express higher-order predicates: tuples in a package 
need to \emph{collectively} satisfy all global constraints. This means that a global constraint is a
property of a package, not of a single tuple. For example, ``{\sf COUNT($*$) $=$
$3$}'' specifies that the entire package should have exactly 3 meals.
\pql also allows the expression of sub-queries in the {\sf SUCH THAT} clause.
In contrast with base constraints, global constraints cannot be expressed by traditional
SQL queries.

The \emph{objective} clause, {\sf MAXIMIZE} (or {\sf MINIMIZE}), is unique to
packages as well: it specifies that out of all packages that satisfy the base
and global constraints, the ones with larger value in the {\sf MAXIMIZE}
clause are preferable.
A detailed description of \pql can be found online~\cite{tech-report}.

\pagebreak
\section{Interface Abstractions}
\label{sec:interface}

Package queries are more complex, semantically and algorithmically, compared
to traditional database queries, and they pose challenges on several fronts:
(a) they can have complex specifications,
(b) they produce a large number of results, which poses usability challenges, and
(c) they are computationally intensive to evaluate.
We discuss the third challenge in Section~\ref{sec:complexity}.
In this section, we describe several interface
abstractions that \pbq implements to address the first two challenges.

\subsection{Specification}
Our \emph{package template} abstraction encodes package specifications in a
familiar tabular format (Figure \ref{fig:template} shows a screenshot example).
The central component of the template is a \emph{sample package}, 
presented as a scrollable table. Additional components
include representations of base and global constraints, optimization
objectives, and suggestions for additional package refinements. As a user
interacts with the template by highlighting elements in the sample package,
\pbq suggests constraints~\cite{wrangler, dataplay}. For example, when the user
selects a cell within the ``fats'' column, the system proposes
several constraints that would restrict the amount of fat in each meal, and
objectives that would minimize the total amount of fat.
The package template is quite expressive but is not as powerful as the \pql
language itself. The abstraction tries to strike a balance between ease-of-use
and expressive power.

\subsection{Presentation}
In addition, \pbq presents packages in a way that allows users to meaningfully 
view the entire package space, without having to actually examine it in its entirety
(see the \emph{visual summary} at the bottom of Figure~\ref{fig:template}).
The system analyzes the current query specification and selects two dimensions to
visually layout the valid packages along. Users can use the visual summary
to navigate through the available packages by selecting glyphs that represent
them.

\subsection{Adaptive exploration}
\label{sec:explore}
Many users may prefer specifying queries in trial-and-error, incremental form,
rather than providing a complete and precise specification from
the very beginning. To facilitate this approach, \pbq initially presents a sample package that
satisfies a few basic constraints. Users can then select good tuples within
the sample, and request a new sample that replaces the unselected tuples. Users can
repeat this process until they reach the ideal package. \pbq uses these
selections to narrow the search space as well as to identify additional
package constraints.

\section{Evaluation}
\label{sec:complexity}

Evaluating package queries is nontrivial: even if the package query does not allow
duplicate tuples, the number of valid packages is in the worst case
exponential in the number of base tuples. In fact, any subset of the base tuples may
potentially be a valid package.
A brute-force approach that generates and evaluates all candidate packages is
thus impractical.

\pbq is an external module which communicates with the DBMS, where the data
resides, via SQL. To evaluate a package query, the system parses a \pql query
and performs a search to generate valid packages. The system either: (i) uses
SQL statements to generate and validate candidate packages; or (ii) translates
package queries to constraint optimization problems, and employs
state-of-the-art constraint solvers to derive valid packages. At the heart of
the query evaluation system, \pbq uses and extends the Tiresias query
engine~\cite{meliou2012tiresias}. Even though \pbq uses the Tiresias query
engine, it has several differences:
\begin{itemize}[noitemsep,leftmargin=0.5cm]
	\item Package queries specify tuple collections (packages), whereas Tiresias' how-to
	queries specify updates to underlying datasets.
	\item \pbq allows a tuple to appear multiple times in a package
	result; this does not map to any operation in Tiresias.
	\item \pql is SQL-based whereas Tiresias uses a variant of Datalog.
	\item \pbq supports arbitrary Boolean formulas in the {\sf SUCH THAT} clause,
	whereas Tiresias only supports conjunctive how-to queries.
	\item \pbq employs additional heuristic and pruning techniques to increase the
	efficiency of package queries.
\end{itemize}
We proceed to describe, at a high level, some of the extensions to Tiresias
used in \pbq to evaluate package queries.

\subsection{Cardinality-based pruning}

With pruning techniques, the system can avoid generating candidate
packages that cannot possibly satisfy some of the global constraints.
Given a global constraint $\mathcal{C}$,
our pruning strategy identifies a lower cardinality bound $l$ and an
upper cardinality bound $u$ for any package that can satisfy $\mathcal{C}$.
For example, if $\mathcal{C}$ is defined as $a \le {\sf COUNT(*)} \le b$, the cardinality bounds 
are trivially $l=a$ and $u=b$.
As another example, consider the global constraint on total calories per package: $2000 \le {\sf
SUM(calories)} \le 2500$. In this case, the cardinality bounds are $l = \lceil \frac{2000}{{\sf MAX(calories)}}
\rceil$ and $u = \lfloor \frac{3000}{{\sf MIN(calories)}} \rfloor$. In fact, with at
least $l$ recipes with {\sf MAX(calories)} and at most $u$ recipes with {\sf
MIN(calories)} we can achieve both the lower and upper bounds of the summation
constraint.

Assuming queries that do not allow repeated tuples,
if $n$ tuples satisfy the base constraints,
this pruning approach reduces the search space from
$2^n$ to $\binom{n}{l} + \binom{n}{l+1} + \cdots + \binom{n}{u-1} + \binom{n}{u}$,
without losing any valid solution.

\subsection{Heuristic local search}

Pruning often reduces the search space significantly, but this reduction alone
is seldom sufficient. In addition to pruning algorithms, which reduce the
search space while maintaining completeness, \pbq employs a heuristic local search
to hasten the computation.
As with any heuristic, there is no guarantee that all valid solutions will be found.
Given a starting package $P_0$ (which can be constructed, for example, at
random), \pbq identifies all possible $k$-tuple replacements that can lead to a valid package,
by using a single SQL query.
For example, suppose we wish to generate meal packages with less than 2,500
total calories. Given a package $P_0$ having a total of 3,000 calories, we
can identify all possible single-tuple replacements which lead to valid packages 
with the following SQL query:
{\sf
\begin{tabbing}
\hspace*{5mm}\=\hspace*{2cm}\=\hspace*{3cm}\= \kill
\>SELECT  \>$P_0$.id, $R$.id\\
\>FROM    \>$P_0$, Recipes $R$\\
\>WHERE   \>$3000$ $-$ $P_0$.calories $+$ $R$.calories $\leq$ $2500$
\end{tabbing}}
This query implements a greedy heuristic that is only able to locate valid packages
that differ from $P_0$ by one single tuple. It fails to find any valid package that
differ from $P_0$ by more than one tuple.
The query can be also modified to explore packages of different cardinalities in 
a straightforward way.
Notice that the query is a selection over a Cartesian product between the candidate
package and the recipe relation. This approach is very efficient if we are
attempting to replace only a few tuples at a time. For $k$ replacements,
however, this method would require a $2k$-way join, which quickly becomes
intractable.

This local search is also particularly useful for adaptive exploration (Section
\ref{sec:explore}), where users usually request the replacement of only a few
tuples at a time.

\section{Challenges}

Package queries pose a series of new challenges on database query engines.
We discuss here a few of the research directions that we plan to explore.

\begin{description}[leftmargin=0mm, topsep=0mm, itemsep=0mm]
\item[Optimizing \pql queries:]
    Our experience with \pbq shows that each of the evaluation techniques we
    adopted have different strengths and weaknesses. Currently, \pbq
    heuristically combines all of them to efficiently derive packages.
    However, a more principled approach to package query optimization could
    add several benefits to the query engine.

\item[Solver limitations:]
    Constraint solvers are typically limited to returning a single package solution at a
    time, and retrieving more packages requires modifying and re-evaluating the query.
    Moreover, solvers cannot usually handle non-linear global constraints;
    hence evaluating such queries requires different methods.

\item[Diverse package results:]
    The number of solutions to a package query can potentially be extremely
    large, making it harder for users to explore the package space and find
    interesting packages. We plan to devise techniques to present the user
    with the most diverse and potentially interesting packages, extracted from
    the space of valid or invalid solutions.
\end{description}

\section{Related Work}
\label{sec:related-work}
Package queries are instances of \emph{constraint satisfaction problems}~\cite{russell1995artificial}, 
a well-known class of NP-complete problems.
Package queries can be used to provide \emph{set-based recommendations}, such
as those available in CourseRank~\cite{course-rank}. \pql offers a more
general framework for package recommendations. For instance, it can easily
express pre-requisite constraints typical of course package recommendation
systems.
\pbq extends the Tiresias query engine~\cite{meliou2012tiresias} with several new features,
which we discussed in Section~\ref{sec:complexity}.

Package queries are also related to skyline
queries~\cite{borzsony2001skyline}, top-$k$ queries~\cite{ilyas2008survey} and
multi-objective queries~\cite{balke2004multi}, in their intent to let users
filter a set of objects based on optimization objectives. However, \pql
queries differ from them for several reasons. First, the optimization
objectives of a skyline query are per tuple, rather than per package. This
makes it possible to express a skyline query with traditional nested
SQL~\cite{borzsony2001skyline}, whereas the global constraints expressible in
\pql are not expressible in traditional SQL. Secondly, \pbq supports
one single per-package optimization objective, as opposed to multiple
per-tuple objectives supported by multi-objective and skyline queries, and
does not support top-$k$ queries. Finally, multi-objective queries (comprising
skyline queries) return the set of non-dominated objects. \pbq, instead,
returns the optimal package for any given query.

\balance

\section{Demo Scenario}
\label{sec:scenario}

VLDB attendees visiting our booth will learn, test, and use \pbq. We will demo
\pbq on a real-world application: the \emph{meal planner}. Meal planner has a rich
recipe data set scrapped from online recipe and nutrition websites. Attendees
will observe how packages are specified with the package template, and
interactively refined with adaptive exploration. In addition, a quick tutorial
will highlight the key features of \pql and describe the query engine
optimizations we employ to optimize the package search process.
For instance, we will show how a \pql query is translated into a linear program
and then solved using existing constraint solvers.
To control
booth crowds, we will provide video tutorials and online guides, and make them
accessible through handheld devices present at the booth. Attendees can choose
to learn about \pbq either by using these materials or by interacting with the
presenters.

Attendees will then test \pbq by building their own recipe packages using
either the visual interface or \pql. Attendees can save their packages, as
well as share their results through tweets or emails. The meal planner will be
accessible online throughout the duration of the conference and users will be able to
revise their saved packages at any time.

\smallskip
\noindent
\textbf{Acknowledgements.}
This paper has appeared as a demonstration in PVLDB 2014. This work was
partially supported by the National Science Foundation under grants
IIS-1421322 and IIS-1420941.

\bibliographystyle{abbrv}
\bibliography{refs}
\balancecolumns

\end{document}